\documentclass[
 aip,
 jcp,
 amsmath,
 amssymb,
]{revtex4-1}
\usepackage[utf8]{inputenc}
\usepackage{array}
\usepackage{amsmath}
\usepackage{float}
\usepackage{xcolor}
\usepackage{graphicx}
\usepackage{braket}

\begin{document}

\title{The variational quantum eigensolver self-consistent field method within a polarizable embedded framework}
\author{Erik Rosendahl Kjellgren}
\email{kjellgren@sdu.dk}
\affiliation{Department of Physics, Chemistry and Pharmacy, University of Southern Denmark, DK-5230 Odense M, Denmark}
\author{Peter Reinholdt}
\affiliation{Department of Physics, Chemistry and Pharmacy, University of Southern Denmark, DK-5230 Odense M, Denmark}
\author{Aaron Fitzpatrick}
\affiliation{Algorithmiq Ltd, Kanavakatu 3C, FI-00160 Helsinki, Finland}
\author{Walter N. Talarico}
\affiliation{Algorithmiq Ltd, Kanavakatu 3C, FI-00160 Helsinki, Finland}
\affiliation{Department of Applied Physics, QTF Centre of Excellence, Center for Quantum
Engineering, Aalto University School of Science, FIN-00076 Aalto, Finland}
\author{Phillip W. K. Jensen}
\affiliation{Department of Chemistry, University of Copenhagen, DK-2100 Copenhagen {\O}, Denmark}
\author{Stephan P. A. Sauer}
\affiliation{Department of Chemistry, University of Copenhagen, DK-2100 Copenhagen {\O}, Denmark}
\author{Sonia Coriani}
\affiliation{DTU Chemistry - Department of Chemistry, Technical University of Denmark, DK-2800 Kongens Lyngby, Denmark}
\author{Stefan Knecht}
\affiliation{Algorithmiq Ltd, Kanavakatu 3C, FI-00160 Helsinki, Finland}
\affiliation{ETH Z{\"u}rich, Department of Chemistry and Applied Life Sciences,
Vladimir-Prelog-Weg 1-5/10, CH-8093 Z{\"u}rich, Switzerland}
\author{Jacob Kongsted}
\affiliation{Department of Physics, Chemistry and Pharmacy, University of Southern Denmark, DK-5230 Odense M, Denmark}

\date{\today}

\begin{abstract}
We formulate and implement the Variational Quantum Eigensolver Self Consistent Field (VQE-SCF) algorithm in combination with polarizable embedding (PE), thereby extending PE to the regime of quantum computing. We test the resulting algorithm, PE-VQE-SCF, on quantum simulators and demonstrate that the computational stress on the quantum device is only slightly increased in terms of gate counts compared to regular VQE-SCF. On the other hand, no increase in shot noise was observed. We illustrate how PE-VQE-SCF may lead to the modeling of real chemical systems using a simulation of the reaction barrier of the Diels-Alder reaction between furan and ethene as an example. 
\end{abstract}


\maketitle

\section{Introduction}
Quantum computers are expected to enable the solution of many computational problems that are intractable on classical high-performance computers. 
The probably most often mentioned example is quantum chemistry, i.e., the solution of the electronic Schr{\"o}dinger equation.~\cite{Troyer2023} 
Quantum chemists have always been very receptive to advances in computer technology to improve the performance of quantum chemistry programs, devising theories and algorithms that could take advantage of the new hardware.~\cite{Gordon2020} 
Not surprisingly, since the first (proof-of-principle) application of a quantum computing algorithm for computational chemistry problems in 2005,~\cite{Aspuru-Guzik2005} the interest of the quantum chemistry community in the development of quantum chemistry methods on quantum processing units has been steadily growing, and almost exploded during the last five years in concomitance with the advances on the hardware site. 
Setting out from different angles and perspectives, a  number of recent comprehensive reviews discuss the current state of affairs concerning the development of theories, methods, and algorithms that aim at performing quantum chemical calculations on both near-term noisy intermediate-scale quantum (NISQ) devices and early/far-future, 
fault-tolerant ones.~\cite{Cao2019,Bauer2020,Elfving2020,McArdle2020b,Motta2022,Aulicino2022,Bharti2022,Liu2022} 
As of today, potential quantum advantage has only been demonstrated for model systems, though.~\cite{Liu2022}

The current paradigm for calculations on NISQ devices is to employ a hybrid quantum-classical approach,~\cite{McClean2016} where only parts of the quantum chemical calculation are actually carried out on the quantum computer, while the remaining parts are run on classical CPU- or GPU-based computers. 
The calculations carried out on the QPUs mostly involve evaluating expectation values over a Hamiltonian or other quantum mechanical operators. 
Often a Variational Quantum Eigensolver (VQE) algorithm~\cite{Yung2014,Cao2019,Romero2019,Wang2019,Fedorov2022b} is employed in this step. 

Within the realm of quantum chemical methods for NISQ devices, the majority of the published work concerning wavefunction ans{\"a}tze focuses on some variants of unitary coupled cluster (UCC) theory.~\cite{McClean2016, Romero2019, Fedorov2022b} 
It is believed~\cite{Romero2019} that UCC methods intrinsically are, or can be made, more easily multi-configurational than the traditional coupled cluster ones.~\cite{Helgaker2013-xk, Christiansen2006, Bartlett2012} 
This is a very important aspect as several interesting chemical problems, especially within life sciences, pertain to large and complicated multiconfigurational systems, the properties and chemistry of which stand little chance of being accurately described by classical computers.~\cite{Elfving2020}  
Approaches like complete-active space self-consistent field (CASSCF),~\cite{Siegbahn1980,Roos1980,Siegbahn1981} complete-active space configuration interaction (CASCI)~\cite{Olsen1988} or density matrix renormalization group (DMRG),~\cite{Baiardi2020} which all involve a full configuration interaction type wavefunction within the space of the chemical relevant orbitals, are typically used to treat molecular systems with multiconfigurational character on classical computers, but quickly hit the ceiling of what can be handled in terms of size of the relevant active space. 
Therefore, exploring the possibility of performing CASSCF/CI calculations using quantum processors is natural.~\cite{Parrish2019, Bauman2021, Fitzpatrick2022} 

However, most of chemistry and all of biochemistry happen in solution. Furthermore, the current most accurate approach to model the reactions catalyzed by enzymes is to treat the active site with some quantum chemical method and describe the protein environment using some molecular mechanics model, i.e., the hybrid QM/MM approach. 
For quantum chemical simulations on classical computers, many methods for treating solvent or environment effects have been developed and implemented, ranging from simple continuum solvent models like the polarizable continuum model (PCM)~\cite{Mennucci2012} to an explicit treatment of solvent molecules or the fragments of molecules in the environment via the Polarizable Embedding (PE) model.~\cite{olsen2010,olsen2011} For calculations on quantum computers, 
on the other hand, only simple approaches such as the PCM method~\cite{Castaldo2022} and point-charge embeddings\cite{hohenstein2023efficient} have so far been implemented in combination with VQE calculations of ground state energies. 
More sophisticated quantum embedding schemes\cite{rossmannek2023quantum} could potentially be applied to include solvation effects, although a practical demonstration has yet to be reported. 
In this work, we present, therefore, an implementation of the PE model for quantum computers in combination with the Adaptive Derivative-Assembled Problem-Tailored Ansatz Variational Quantum Eigensolver self-consistent field approach (ADAPT-VQE-SCF),~\cite{Fitzpatrick2022} and illustrate its performance for ammonia in water and for the Diels-Alder reaction of furan with ethene in water.

\section{Theoretical summary}

In the following, we outline the theoretical background for our novel developments. 
This theoretical summary is split into three parts:
first, we introduce the polarizable embedding model; second, the working equations from the variational quantum eigensolver self-consistent field method are briefly discussed; and third, we show how to integrate the PE model with the ADAPT-VQE-SCF scheme. 

\subsection{Polarizable Embedding}
\begin{figure}
    \centering
    \includegraphics{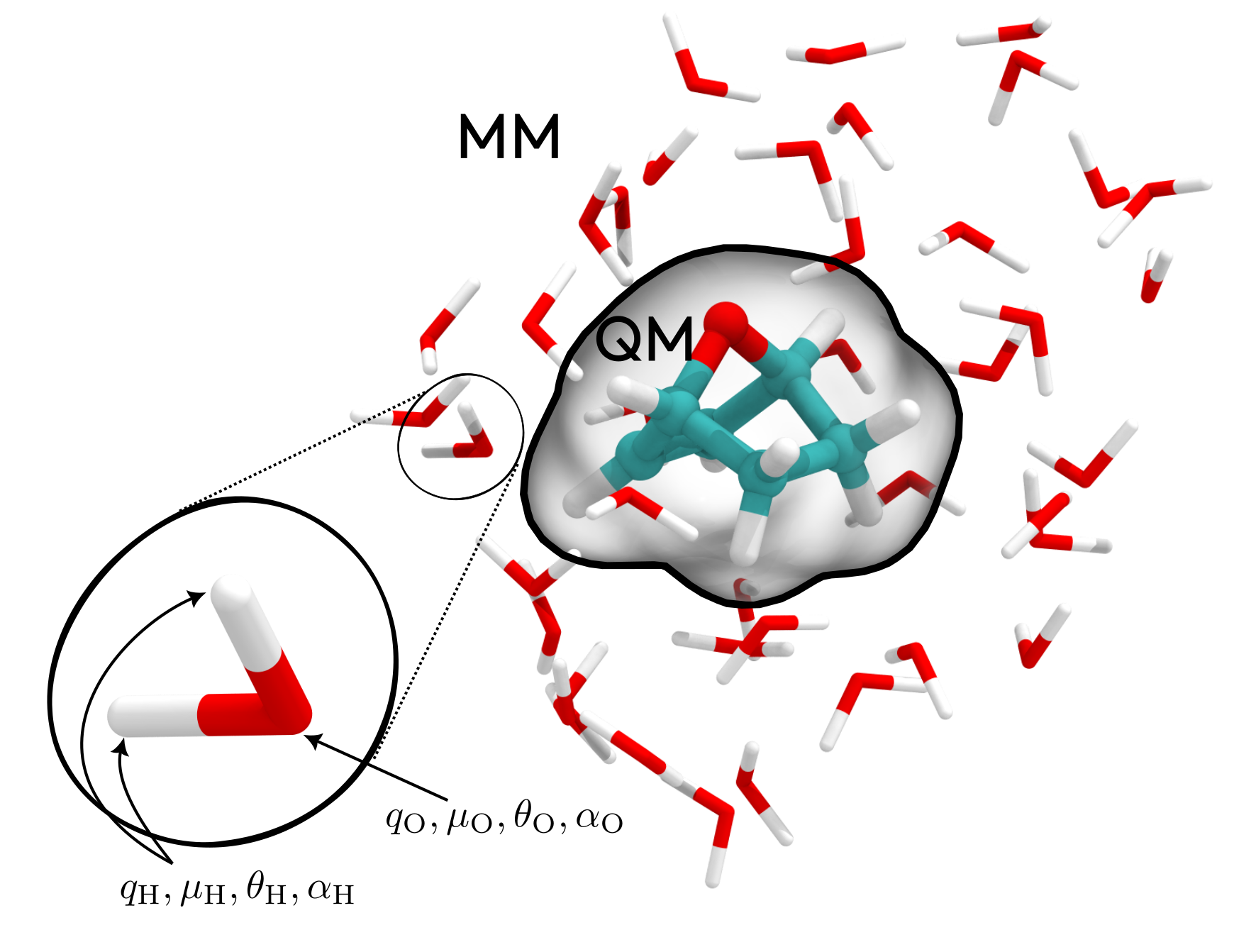}
    \caption{The QM/MM partitioning in the PE model. A complete system is divided into QM and MM subsystems, where the QM part is described through an accurate quantum chemical method. The environment is divided into smaller fragments, described by an atomic multi-center multipole expansion and atomic polarizabilities.}
    \label{fig:pe-illustration}
\end{figure}

The Polarizable Embedding\cite{olsen2010,olsen2011} (PE) model divides a total molecular system into a smaller quantum region and an environment. The quantum region is described using an accurate quantum chemistry model, whereas the environment is represented semi-classically through multipoles and polarizabilities. The parameters describing the embedding potential are generated by dividing the environment into small fragments and carrying out individual quantum-chemical calculations for each of them to obtain atomic-based multipoles and polarizabilities using a multi-center multipole expansion. The multipoles and polarizabilities represent the permanent and induced charge density of the fragments in the environment, respectively. 
The partitioning of the system in QM and polarizable MM regions is graphically illustrated in Figure \ref{fig:pe-illustration}.

The effects of the environment on the quantum region are included by constructing an effective Hamiltonian
\begin{equation} \label{eq:effectivefock}
\hat{H}^{\mathrm{eff}}=\hat{H}^{\mathrm{vac}}+\hat{v}^{\mathrm{PE}} 
\end{equation}
where $\hat{H}^{\mathrm{vac}}$ represents the regular (isolated) Hamiltonian for the quantum region without the presence of the environment. The operator $\hat{v}^{\mathrm{PE}}$ is the PE operator, which describes the potential from the environment. The PE operator is further divided into an electrostatic and an induction (polarization) term
\begin{equation} \label{eq:v_PE}
\hat{v}^{\mathrm{PE}}=\hat{v}^{\mathrm{es}}+\hat{v}^{\mathrm{ind}} 
\end{equation}
where the electrostatic operator, $\hat{v}^{\mathrm{es}}$, describes the potential from the permanent charge distributions of the $N$ environment sites
\begin{equation} \label{eq:v_es}
\hat{v}^{\mathrm{es}}=\sum_{s=1}^{N}\sum_{\left|k\right|=0}^{K}\frac{\left(-1\right)^{\left|k\right|}}{k!}M_{s}^{\left(k\right)}\hat{V}_{s,\mathrm{el}}^{\left(k\right)}
\end{equation}
written using a multi-index notation,\cite{olsen2014thesis} i.e., $k$ runs over the Cartesian components of the multipole moment tensors.
Here, $M_{s}^{\left(k\right)}$ are the multipoles on expansion site $s$, i.e., it is the $k$'th components of the atomic charge, dipole and quadrupole moment of site $s$.
The operator $\hat{V}_{s,\mathrm{el}}^{\left(k\right)}$ gives a derivative of the electric potential at site $s$, and is defined as
\begin{equation}
    \hat{V}_{s,\mathrm{el}}^{(k)} = \sum_{pq}   t^{(k)}_{pq} (\mathbf{R}_s) \hat{E}_{pq}
\end{equation}
where $\hat{E}_{pq}=\hat{a}^\dagger_{p\alpha}\hat{a}_{q\alpha}+\hat{a}^\dagger_{p\beta}\hat{a}_{q\beta}$ is the singlet one-electron excitation
operator,\cite{Helgaker2013-xk} and the potential derivative integrals are
\begin{equation}
    t^{(k)}_{pq} (\mathbf{R}_s) = \int \phi^{*}_{p}(\mathbf{r})\nabla^k \left(\frac{1}{\left| \mathbf{r} - \mathbf{R}_s \right|}\right) \phi_{q}(\mathbf{r}) \mathrm{d}\mathbf{r}~.
\end{equation}

The induction operator, $\hat{v}^{\mathrm{ind}}$, originates from the polarized charge distribution of the environment, described by induced dipoles, and it is defined as
\begin{equation} \label{eq:v_ind}
\hat{v}^{\mathrm{ind}}=-\sum_{s=1}^{N}\boldsymbol{\mu}_{s}^{\mathrm{ind}}\left({\mathbf{F}}_{\mathrm{tot}}\right)\hat{V}^{(1)}_{s,\mathrm{el}} \ .
\end{equation}
The induced dipoles $\boldsymbol{\mu}_{s}^{\mathrm{ind}}$ are generated in response to the total electric field ($\mathbf{F}_{\mathrm{tot}}$) at the polarizable site $s$, which is the sum of fields from the electrons and nuclei in the quantum region, the permanent multipole moments in the environment, as well as other induced dipoles in the environment. The operator\ $\hat{{V}}^{(1)}_{s,\mathrm{el}}$ gives the electric field from the electrons at a polarizable site $s$. The induced dipoles entering the expression for the induction operator are derived from 
\begin{equation} \label{eq:indmom}
\boldsymbol{\mathbf{\mu}}_{s}^{\mathrm{ind}}\left(\mathbf{F}_{\mathrm{tot}}\right)=\boldsymbol{\alpha}_{s}\mathbf{F}_{\mathrm{tot}}(\mathbf{R}_{s})=\boldsymbol{\alpha}_{s}\left(\mathbf{F}_{}(\mathbf{R}_{s})+\sum_{s'\neq s}\mathbf{T}_{ss'}^{(2)}\boldsymbol{\mathbf{\mu}}_{s'}^{\mathrm{ind}}\right) \ ,
\end{equation}
where $\mathbf{F}_{}(\mathbf{R}_{s})$ is the electric field at site $s$ from the nuclei, electrons, and permanent multipole moments, and $\mathbf{T}_{ss'}^{(2)}$ is a dipole--dipole interaction tensor.\cite{Stone2013Theory}
Equation~\eqref{eq:indmom} leads to a set of coupled equations which can be formulated as a linear system of equations by introducing a column vector containing the induced dipoles $\boldsymbol{\mu}^{\mathrm{ind}}=\left(\boldsymbol{\mu}_{1}^{\mathrm{ind}},\boldsymbol{\mu}_{2}^{\mathrm{ind}},\ldots,\boldsymbol{\mu}_{N}^{\mathrm{ind}}\right)^{T}$,
and one containing the electric fields $\mathbf{F}=\left(\mathbf{F}_{}(\mathbf{R}_{1}),\mathbf{F}_{}(\mathbf{R}_{2}),\ldots,\mathbf{F}_{}(\mathbf{R}_{N})\right)^{T}$.
The induced dipole moments may formally be obtained as the solution of 
\begin{equation} \label{eq:direct_solver}
\boldsymbol{\mu}^{\mathrm{ind}}=\mathbf{BF} \ ,
\end{equation}
where $\mathbf{B}$ is the ($3N \times 3N$) classical response matrix
\begin{equation} \label{eq:classicalresponsematrix}
\mathbf{B}=\begin{pmatrix}\pmb{\alpha}_{1}^{-1} & -\mathbf{T}_{12}^{(2)} & \ldots & -\mathbf{T}_{1N}^{(2)}\\
-\mathbf{T}_{21}^{(2)} & \pmb{\alpha}_{2}^{-1} & \ddots & \vdots\\
\vdots & \ddots & \ddots & -\mathbf{T}_{(N-1)N}^{(2)}\\
-\mathbf{T}_{N1}^{(2)} & \dots & -\mathbf{T}_{N(N-1)}^{(2)} & \pmb{\alpha}_{N}^{-1}
\end{pmatrix}^{-1} \ .
\end{equation}
In practice, the induced dipoles can be obtained by either explicitly constructing the response matrix and directly solving Equation~\eqref{eq:direct_solver} by conventional means or by iterative schemes when an explicit construction is not computationally feasible due to the potentially large dimensions of the environment.
The induced dipoles depend on the fields exerted by the electron density of the quantum region, which in turn depends on the induced dipoles through the induction operator, and it is thus required to use a self-consistent scheme in which the induced dipoles are updated in every quantum mechanically self-consistent-field cycle. Such a scheme finally leads to a mutual relaxation of the quantum 
wavefunction/density and the induced dipoles of the environment.

\subsection{Variational Quantum Eigensolver Self-Consistent Field}

The following account of the variational quantum eigensolver self-consistent field (VQE-SCF) presents the most important expressions from Fitzpatrick et al.\cite{Fitzpatrick2022} The molecular spin-free full (non-relativistic) electronic Hamiltonian, $\hat{H}$, is the starting point for the VQE-SCF approach
\begin{equation}
    \hat{H} = \sum_{pq}h_{pq}\hat{E}_{pq} + \frac{1}{2}\sum_{pqrs}g_{pqrs}\left(\hat{E}_{pq}\hat{E}_{rs}-\delta_{qr}\hat{E}_{ps}\right)
    \label{eq:mol_H}
\end{equation}
with the one- and two-electron integrals, respectively,
\begin{equation}
    h_{pq} = \int \phi^*_p\left(\mathbf{r}\right)\hat{h} \ \phi_q\left(\mathbf{r}\right) \mathrm{d}\mathbf{r}
\end{equation}
\begin{equation}
    g_{pqrs} = \int \phi^*_p\left(\mathbf{r}_1\right)\phi^*_r\left(\mathbf{r}_2\right)r^{-1}_{12}\phi_q\left(\mathbf{r}_1\right)\phi_s\left(\mathbf{r}_2\right) \mathrm{d}\mathbf{r}_1\mathrm{d}\mathbf{r}_2~.
\end{equation}
The set of functions $\phi_p$ represents an orthonormal molecular orbital (MO) basis,  i.e., $\left<\phi_p\left|\phi_q\right.\right>=\delta_{pq}$. 
In the framework of VQE-SCF, the wavefunction is parameterized by orbital rotation coefficients, $\kappa_{pq}$, and unitary qubit rotation angles, $\theta_i$.
The parameterized wavefunction takes the form,
\begin{equation}
    \left|\Psi\left(\boldsymbol{\kappa}, \boldsymbol{\theta}\right)\right> = U\left(\boldsymbol{\kappa}\right)U\left(\boldsymbol{\theta}\right)\left|0\right>
\end{equation}
with $\left|0\right>$ being the reference state.
The orbital rotation parameterization takes the same form as in the conventional multi-configurational self-consistent field (MCSCF) methods,~\cite{Helgaker2013-xk}
\begin{equation}
    U\left(\boldsymbol{\kappa}\right) = \exp\left({\hat \kappa}\right)
\end{equation}
with the anti-hermitian orbital rotation operator being
\begin{equation}
    {\hat \kappa} = \sum_{p>q}\kappa_{pq}\left(\hat{E}_{pq}-\hat{E}_{qp}\right)~.
\end{equation}
The parameterization of the unitary qubit rotation angles is where the VQE-SCF differs from the conventional MCSCF.
In conventional MCSCF, the wavefunction is usually parameterized with configuration interaction (CI) linear-combination coefficients or, in unitary form, by the state-transfer operator.
Instead, to target a quantum computer, the parameterization uses unitary qubit rotation angles.
This parameterization takes the form
\begin{equation}
    U\left(\boldsymbol{\theta}\right) = \prod_i\exp\left(i\theta_i \hat P_i\right)
\end{equation}
with $\hat P_i$ being strings of Pauli operators.
The ground state energy is determined by minimization with respect to both the orbital rotation coefficients and the unitary qubit rotation angles, 
\begin{equation}
    E = \min_{\boldsymbol{\theta},\boldsymbol{\kappa}}\left<0\left|U^\dagger\left(\boldsymbol{\theta}\right)U^\dagger\left(\boldsymbol{\kappa}\right)\hat{H} U\left(\boldsymbol{\kappa}\right)U\left(\boldsymbol{\theta}\right)\right|0\right>~.
\end{equation}
The choice of specific unitary qubit rotation parameters in the ansatz follows the ADAPT approach and can be either ADAPT-VQE~\cite{grimsley2019adaptive} or qubit-ADAPT-VQE.\cite{tang2021qubit}

\subsection{PE-VQE-SCF}

The following briefly outlines how to integrate the VQE approach with PE. In the PE-VQE-SCF model, instead of minimizing the energy with respect to the molecular Hamiltonian, as in Eq.~\eqref{eq:mol_H}, we minimize it with respect to the free-energy in solution  
\begin{equation}
    E_\text{PE} = \min_{\boldsymbol{\theta},\boldsymbol{\kappa}}\left<0\left|U^\dagger\left(\boldsymbol{\theta}\right)U^\dagger\left(\boldsymbol{\kappa}\right)\hat{H}^\text{PE} U\left(\boldsymbol{\kappa}\right)U\left(\boldsymbol{\theta}\right)\right|0\right>
\end{equation}
where the Hamiltonian is defined as 
\begin{equation}
    \hat{H}^\text{PE} = \hat{H}^{\mathrm{eff}} - \frac{1}{2}\hat{v}^{\mathrm{ind}}.
\end{equation}
For the VQE-SCF algorithm, the effective Hamiltonian needs to be decomposed into Pauli strings.
The underlying fermionic operators in the operators $\hat{V}_{s,\mathrm{el}}^{\left(k\right)}$ and $\hat{\mathbf{F}}_{s,\mathrm{el}}$, 
Eq.~\eqref{eq:v_es} and Eq.~\eqref{eq:v_ind}, respectively, are the singlet single excitation operators $\hat{E}_{pq}$.
In the gas-phase, this corresponds to the following decomposition of the Hamiltonian
\begin{equation}
    \hat{H}^\text{vac}(\boldsymbol{\kappa}) = \sum_i c_i(\boldsymbol{\kappa})\hat{P}_i~, \label{eq:decomp-vac}
\end{equation}
where $c_i(\kappa)$ is the weight of the Pauli strings from the decomposition and $\hat{P}_i$ is a unique Pauli string.
After introducing the PE environment, the decomposition of the Hamiltonian becomes
\begin{equation}
    \hat{H}^\text{PE}\left(\boldsymbol{\kappa}, \mathbf{F}_\text{tot}\right) = \sum_i \tilde{c}_i(\boldsymbol{\kappa}, \mathbf{F}_\text{tot})\hat{P}_i,
\end{equation}
which is similar to eq. \eqref{eq:decomp-vac}, but with modified expansion coefficients $\tilde{c}_i$. Crucially, these weights depend on electric fields from the polarizable environment instead of just the orbital rotation parameters.

Due to the induced dipoles in the environment, the PE Hamiltonian is non-linear in the 
wavefunction parametrization. This leads to a double-SCF algorithm where the wavefunction parameters and the induced dipoles are iteratively minimized, thereby leading to a computational overhead within PE compared to calculating the corresponding molecule in a vacuum.

\begin{figure}[H]
    \centering
    \includegraphics[width=1.0\textwidth]{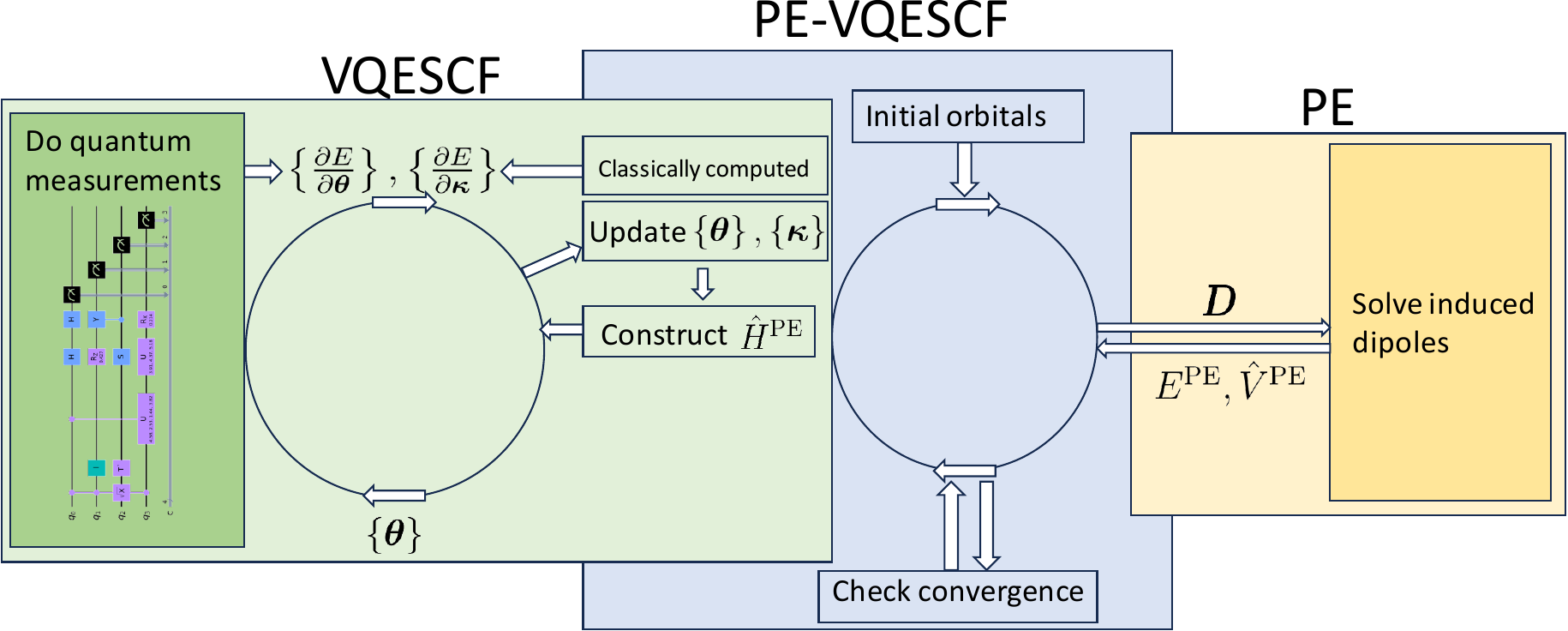}
    \caption{Flow chart of the PE-VQE-SCF algorithm. In this flow chart $D$ is the one-electron reduced density matrix.}
    \label{fig:pevqescf_flow}
\end{figure}

The algorithmic approach underlying the PE-VQE-SCF methods is illustrated in Figure~\ref{fig:pevqescf_flow}.
In this figure, the overall structure of the PE-VQ-SCF algorithm can be seen.
The blue box shows the SCF loop that holds together the PE model and the VQE-SCF procedure.
The yellow box shows what the PE library handles, with input and output.
Finally, the green box illustrates the VQE-SCF algorithm.
It can be seen that the PE library provides the PE contribution to the Hamiltonian, which is then used to solve the VQE-SCF problem.
The VQE-SCF procedure provides a density matrix to the PE library that is needed to solve for the PE contribution (the induced dipoles).
This cycle is continued until convergence is achieved.

\section{Computational Details}
The VQE-SCF calculations were run using \textit{AURORA},\cite{aurora} with PySCF\cite{Sun2017, Sun2020} as the backend for the orbital optimization and CPPE\cite{scheurer2019cppe} as the induced dipole moment solver for the PE model. For the ammonia systems (as described below), ADAPT-VQE was used as the ansatz for the VQE-SCF calculations.
The VQE-SCF calculations were optimized according to the algorithm described in Fitzpatrick et al.\cite{Fitzpatrick2022}
Furthermore, to simulate the probabilistic nature of the measurement outcome of a quantum computer, we used Adaptive Informationally complete generalized Measurements (AIM) implemented as positive operator valued measurements (IC-POVMs)\cite{garcia2021learning, glos22a} from the \textit{AURORA} package. 
As we use informationally complete POVM for the estimation of physical observables $\Bar{O}$ with a finite set of measurement outcomes (very much like Monte Carlo sampling), the estimation of these observables comes with a statistical error originating from the finite number of samples $S$ considered and has the expected scaling $\sigma_{\Bar{O}} \sim \frac{1}{\sqrt{S}}$.
This scheme not only allows us to simulate the shot noise of the measurements but also to convert the AIM outcomes into estimates of different observables\cite{Fitzpatrick2022, nyka22a, filippov2023scalable} needed for the proposed PE-VQE-SCF method.

All calculations performed on the ammonia systems utilized the aug-cc-pVTZ\cite{dunning1989a, kendall1992a} basis set for the quantum region (ammonia), and the loprop-aug-cc-pVTZ\cite{gagliardi2004local} basis set for the calculation of the atomic multipole moments and atomic polarizabilities within the PE region.
The active space comprised 6 electrons in 6 spatial orbitals, and was picked on the basis of MP2 natural orbital occupation numbers.

To illustrate the PE-VQE-SCF approach further, the reaction profile for the Diels-Alder reaction between furan and ethene (reactant) to form (1R,4S)-7-oxabicyclo[2.2.1]hept-2-ene (product) was simulated using the nudged elastic band (NEB) method.\cite{henkelman2000climbing,goumans2009embedded}
The reactant and product structures were placed in a sphere of 40 water molecules using the Packmol\cite{martinez2009packmol} program. To ensure that the two structures were as similar as possible for the following NEB calculation, the geometry of the two structures was minimized using CAM-B3LYP\cite{yanai2004new}-D3BJ\cite{grimme2010consistent,grimme2011effect}/6-31G*\cite{ditchfield1971self,hehre1972self,hariharan1973influence} using Terachem\cite{ufimtsev2009quantum,titov2013generating,seritan2021terachem} (version 1.95A) for energies and gradients with shared solvent coordinates, i.e., we minimize the sum of the energies of the solvated reactant and product structures with respect to the coordinates of the reactant, the coordinates of the product, and the coordinates of the water (shared between the solvated reactant/product structures). The minimization was done in cartesian coordinates with a custom Python script using SLSQP\cite{kraft1988software} from SciPy.\cite{virtanen2020scipy}
The NEB path was generated from these starting structures with Terachem with CAM-B3LYP-D3BJ/6-31G*. We used a "free-end" NEB with 25 NEB images and an NEB spring constant of 0.01 a.u..
The (PE-)VQE-SCF calculations of the reaction profile were performed using the 6-31G* basis set and an active space consisting of 6 electrons in 6 spatial orbitals, and was picked on the basis of MP2 natural orbital occupation numbers. 
The ADAPT approach was used in all molecular examples to find a suitable parametrized quantum circuit describing the multiconfigurational wavefunction within the considered active spaces.

\section{Results}
The result section is structured into two parts. In part A, we aim to test the implemented polarizable embedding strategy in relation to regular VQE-SCF. Our focus in this respect will be on important parameters regarding the efficiency of quantum devices, e.g., gate count and simulated shot noise. For this investigation, we will use a chemical model system consisting of ammonia solvated in water, and we will monitor the above-mentioned parameters as a function of the number of water molecules included in the calculation. All water molecules will be described using the polarizable embedding potential, whereas ammonia will be considered at the QM level of theory. In part B, we illustrate the PE-VQE-SCF approach for the simulation of a reaction coordinate profile of a Diels-Alder reaction, taking directly into account the effects of solvation. We include all the solvating water molecules in the polarizable embedding description. 
We note that we use a limited number of snapshots in the PE calculations in these examples. In general, statistical sampling of the solvent configurations, for example, sampled from a molecular dynamics trajectory, is needed for realistic computational applications. 

\subsection{Ammonia}
\label{PartA}
In this section, we will consider ammonia using PE-VQE-SCF as a test system to access key performance parameters related to the VQE-SCF algorithm. 
\begin{figure}[H]
    \centering
    \includegraphics{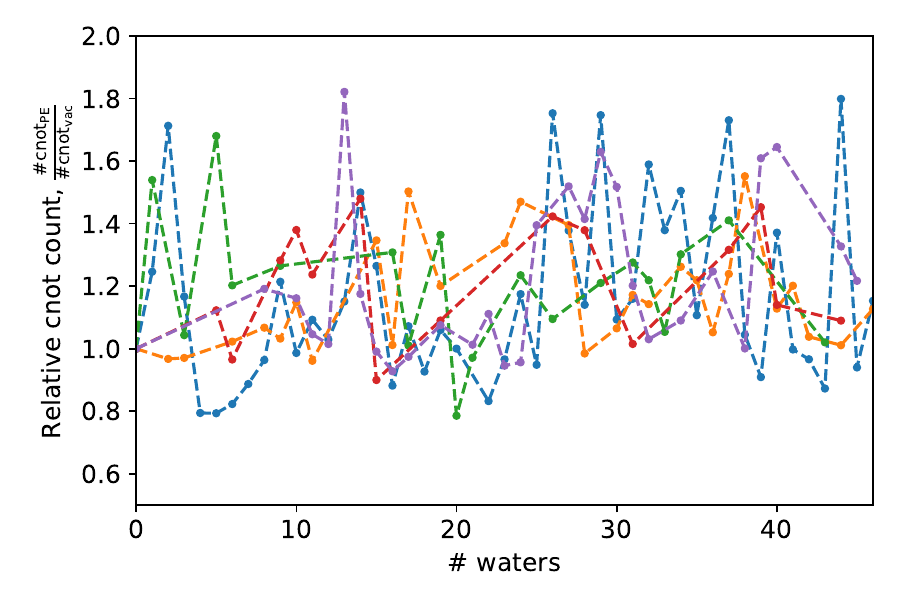}
    \caption{Relative CNOT count with respect to ammonia in isolation (no polarizable embedding) as a function of the number of water molecules in the polarizable embedding potential.
    The different colors represent different snapshots of water configurations.}
    \label{fig:ammonia_cnots}
\end{figure}
In Figure~\ref{fig:ammonia_cnots}, the number of required CNOT gates as a function of the size of the PE region (increasing number of water molecules) relative to the corresponding energy calculation of ammonia in isolation is shown. Five different snapshots have been considered in order to avoid any bias towards just a single solvent configuration. 
From Figure~\ref{fig:ammonia_cnots}, we observe that adding the PE model to the system, on average, will slightly increase the gate count of the calculation compared to the corresponding calculation of ammonia in isolation. On average, the gate count is increased by 16\% when using the PE model for this system. Using ADAPT-VQE, it is to be expected that a larger gate count will appear when including the environment in the system since the wave function will be more complicated (e.g., lack of point-group symmetry) than the wave function describing the system in isolation.
However, as further observed from Figure \ref{fig:ammonia_cnots}, we do not observe a systematic increase in the CNOT count when increasing the size of the embedding part of the system, and the increase in the number of water molecules does not lead, on this basis, to a systematic upward trend regarding the complexity of the underlying circuit.

\begin{figure}[H]
    \centering
    \includegraphics{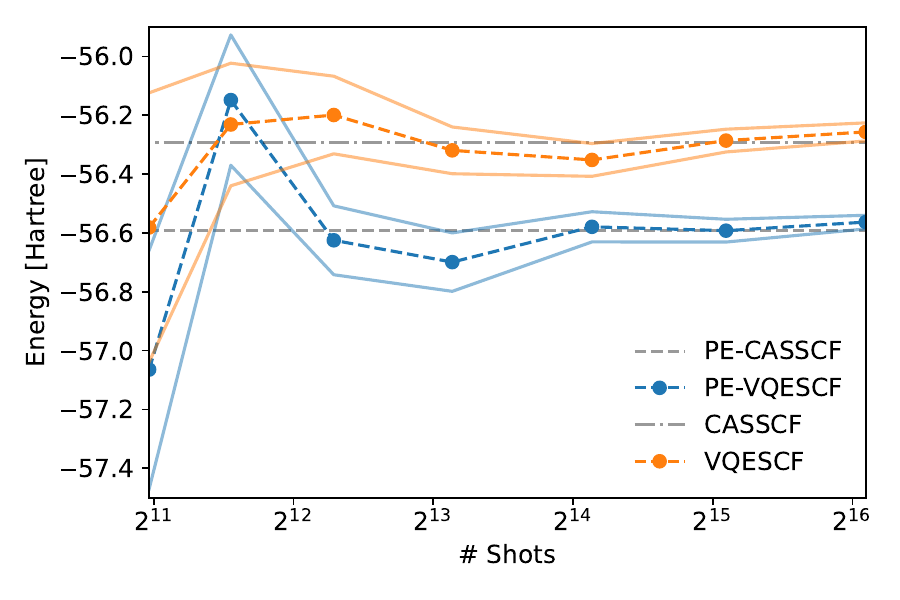}
    \caption{Mean energy and shot noise indicated as plus/minus the standard error. Dashed curves represent the energy, while the full lines represent the error interval. Orange color refers to VQE-SCF while blue color represents PE-VQE-SCF. In total 46 water molecules were included in the PE region.}
    \label{fig:ammonia_shotnoise}
\end{figure}

In Figure \ref{fig:ammonia_shotnoise}, we inspect the mean energy and the simulated shot noise as a function of the number of sampling shots for either VQE-SCF or PE-VQE-SCF.
It is immediately clear that the energy of the PE-VQE-SCF calculations is lower than the corresponding energy of the VQE-SCF calculation. This is expected since the PE model will add terms to the Hamiltonian that will decrease the total energy of the system, i.e., the solvating water will stabilize the ammonia molecule. As expected, the sampling error reduces with an increasing number of shots for either VQE-SCF or PE-VQE-SCF, and the solutions converge towards the classically computed CASSCF and PE-CASSCF results. It is further observed that the noise decreases roughly at the same rate for VQE-SCF and PE-VQE-SCF upon increasing the number of shots. On this basis, it is concluded that adding the embedding term to the molecular Hamiltonian does not negatively affect how many shots are needed to obtain reliable energies within the PE-VQE-SCF framework.

\subsection{Reaction curve}
Next, we consider in this section the ability of the PE-VQE-SCF approach to model real applications showcased based on the Diels-Alder reaction between furan and ethene.

\begin{figure}[H]
    \centering
    \includegraphics{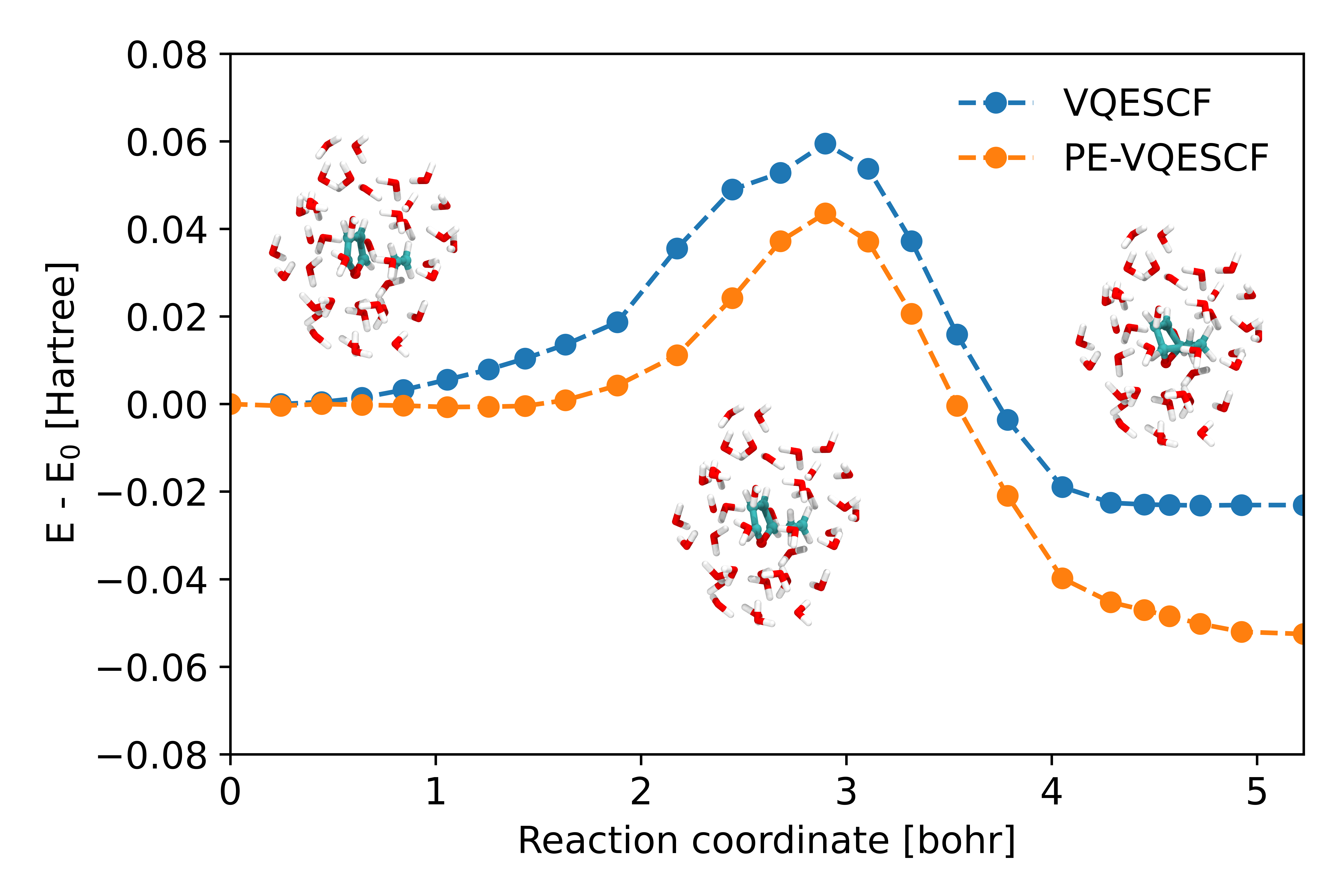}
    \caption{Calculated reaction profile of the Diels-Alder reaction between furan and ethene based on either VQE-SCF (blue line) or PE-VQE-SCF (orange line).
    }
    \label{fig:reaction_curve}
\end{figure}
In Figure \ref{fig:reaction_curve}, the reaction profile of the furan-ethene Diels-Alder reaction is calculated based on either VQE-SCF or PE-VQE-SCF and is shown with respect to the reactant energies obtained in either VQE-SCF or PE-VQE-SCF, i.e., $E-E_0$. In these calculations, furan and ethene are included in the QM part of the system, while all water molecules are described using the embedding potential. 
From Figure \ref{fig:reaction_curve}, we observe a decrease in the calculated barrier upon inclusion of the water solvent, and further, we observe a stabilization of the products. Both these observations, i.e., reduction of the reaction barrier and stabilization of the products, are caused mainly by dipolar stabilization of the chemical reactive part of the system due to the presence of the polar (water) environment. The specific reduction of the reaction barrier is around 0.015 au, corresponding to 9.4 kcal/mol, and the presence of the water environment thereby represents an important part of the chemical system.

\section{Conclusion}
In this work, we have formulated and implemented the VQE-SCF algorithm in combination with polarizable embedding, thereby enabling hybrid quantum-classical simulations on quantum devices. The resulting algorithm, PE-VQE-SCF, has been tested on simulators, where we show that the computational stress on the quantum device is not significantly increased, neither in terms of gate count nor in terms of additional shot noise, upon extending VQE-SCF to PE-VQE-SCF. We have further showcased our PE-VQE-SCF implementation by calculating the Diels-Alder reaction barrier between furan and ethene. We conclude that PE-VQE-SCF represents a robust computational quantum computing strategy for modeling larger and extended chemical systems. Further work will address the ability to use PE-VQE-SCF to tackle biological challenges and extend the computational framework to excited states and molecular property calculations. 

\acknowledgments
We acknowledge the support of the Novo Nordisk Foundation (NNF) for the focused research project ``Hybrid Quantum Chemistry on Hybrid Quantum Computers'' (grant number  NNFSA220080996).

\section*{DATA AVAILABILITY}
The data that support the findings of this study are openly available in Zenodo,\linebreak at https://zenodo.org/records/10279684, reference number 10.5281/zenodo.10279684. 

\bibliography{literature}

\end{document}